\documentclass[iop]{emulateapj}
\usepackage{natbib}
\usepackage{epstopdf}
\shorttitle{Mineralogy and Structure of $\beta$ Pic Inner Disk}
\shortauthors{Li et al.}

\begin{document}

\submitted{Accepted by ApJ}

\title{The Mineralogy and Structure of the Inner Debris Disk of \\$\beta$ Pictoris}

\author{Dan Li, Charles M. Telesco}
\affil{Department of Astronomy, University of Florida,
    Gainesville, FL 32611, USA}

\author{Christopher M. Wright}
\affil{School of Physical, Environmental and Mathematical Sciences, The University of New South Wales, PO Box 7916, Canberra BC ACT 2610, Australia}
\email{dli@astro.ufl.edu}

\begin{abstract}
We observed the edge-on, planet-bearing disk of $\beta$ Pictoris using T-ReCS at Gemini to clarify and extend previous observations and conclusions about this unique system. Our spectroscopy and spectral modeling of the 10-$\micron$ silicate feature constrain the spatial distributions of three representative dust components (0.1-$\micron$/2.0-$\micron$ glassy olivine and crystalline forsterite) across the inner 20-AU of the disk. We confirm that the 2.0-$\micron$ glassy olivine is strongly peaked in the disk center and that the 0.1-$\micron$ glassy olivine does not show this concentration, but rather is double peaked, with the peaks on either side of the star. However, we do not see the strong difference in brightness between those two peaks reported in a previous study. Although the spatial distribution of the 0.1-$\micron$ dust is consistent with the scenario of a dust-replenishing planetesimal belt embedded in the disk, we note an alternative interpretation that can explain the spatial distributions of the 0.1-$\micron$ and 2.0-$\micron$ grains simultaneously and does not require the planetesimal belt. In addition to the spectroscopy, we also obtained a new 11.7 $\micron$ image of the $\beta$ Pic disk. By comparing this image with that acquired in 2003, we confirm the existence and overall shape of the dusty clump at 52 AU in the SW disk. We speculate that the clump's projected spatial displacement of $\sim$2.0 AU, a 3.6-$\sigma$ result, between two epochs separated by seven years is due to the Keplerian motion of the clump at an orbital radius of $54.3^{+2.0}_{-1.2}$ AU. 
\end{abstract}


\section{Introduction}

The infrared emission of a debris disk arises from dust particles generated by collisional cascades initiated by large bodies like planetesimals and comets. For a small number of relatively nearby debris disks, it is becoming possible with 8-10-m-class telescopes to explore directly the relationship among these particles, their parent bodies, and the broader planetary cohort within a system. A particularly spectacular and extensively-studied example is $\beta$ Pictoris, one of the archetypal debris-disk stars with which IRAS originally established this important class. 

$\beta$ Pic is a young (12 Myr, \citealt{zuckerman2001}), A5 V star at a distance of 19.28 pc \citep{crifo1997}. The edge-on disk surrounding it shows a remarkable two-component structure: a main disk and a secondary disk that is inclined by $\sim$5$\degr$ from the main disk and gives rise to the inner disk warp seen in early, low-resolution observations \citep{mouillet1997,heap2000, golimowski2006}. Although the main disk is more extended and brighter in the near-IR, the secondary disk contributes most of the mid-IR flux and exhibits a conspicuous NE/SW asymmetry \citep{lagage1994, pantin1997} and clumpy structures \citep[][here after T5]{telesco2005}. Mid-IR observations also reveal a central zone (r$<$40 AU) where dust is highly depleted \citep{pantin1997}. 10-$\micron$-band spectroscopy of $\beta$ Pic shows the silicate feature similar to that of comet Halley, which suggests a cometary origin of the dust in the inner disk of $\beta$ Pic \citep{telesco1991, knacke1993, aitken1993}. This conclusion is supported by the modeling by \citet{li1998} and observations at other mid-IR wavelengths \citep{chen2007}.  However, collisional processes among km-sized planetesimals may also account for the observed dust properties \citep{thebault2003, grigorieva2007}.

Given the wealth of observational constraints, considerable effort has been expended to produce a coherent picture of the $\beta$ Pic system and properties of its dust \citep[e.g.,][]{pantin1997, li1998, thebault2003, grigorieva2007, czechowski2007, ahmic2009}. It is generally agreed that dust grains are being produced continuously by comets orbiting close to the star, collisional cascades, and/or the collisional break-up of planetesimals, with the particle evolution being strongly affected by the stellar radiation and the gravity of planets in the disk.

Some features of the disk of $\beta$ Pic, such as the inclined secondary disk and its asymmetric morphology, suggest the presence of a planet; in fact, one was discovered recently by direct imaging \citep{lagrange2009, lagrange2010, bonnefoy2011}. The most recent and precise measurements of the position angles (PA) of the main and secondary disks and $\beta$ Pic b imply that the planet is located in the secondary disk \citep{lagrange2012, chauvin2012}, thus providing a promising explanation for its origin. However, at present, other possibilities \citep[e.g.,][]{grigorieva2007} cannot be excluded, and new clues may illuminate the relationship between the planetary system of $\beta$ Pic and the observed disk morphology. 

10-$\micron$-band spectroscopic observations by \citet[hereafter O4]{okamoto2004} using COMICS at the Subaru Telescope reveal what may be several planetesimal belts in the $\beta$ Pic disk, the most prominent of which is located at 6.4 AU from the star and possibly tilted with respect to the secondary disk. T5 observed $\beta$ Pic using T-ReCS \citep{telesco1998} at Gemini South in 2003, and their multi-wavelength images showed a bright clump in the disk at about $\sim$52 AU southwest of the star and composed of particles that may differ in size and/or composition from dust elsewhere in the secondary disk. This bright clump of thermally emitting dust may mark the location of collisionally grinding, resonantly trapped planetesimals or, as T5 emphasized, the cataclysmic break-up of a planetesimal.

To clarify and extend some of these previous observations and conclusions about the $\beta$ Pic disk, we present in this paper new 8-13 $\micron$ spectra and 11.6 $\micron$ (Si-5) images made with sub-arcsec angular resolution across the 40-AU-diameter and the 160-AU-diameter regions in the spectroscopy and imaging, respectively. While the spectra by O4 were obtained at high airmass ($>$3.0) and suffered from strong atmospheric ozone absorption near 9.8 $\micron$, our new data, made recently using T-ReCS at Gemini South, were obtained under more favorable conditions and so deal with some ambiguities in the O4 observations. Using T-ReCS, we also repeated a subset of observations by T5. We use these two-epoch observations, both made with T-ReCS and spanning a period of $\sim$7 years, to constrain further the properties of the bright clump in the disk of $\beta$ Pic. 

The details of our observations and data reduction are presented in the following section. The analysis and modeling of the data are covered in the Section 3. Finally, Sections 4 and 5 are devoted to discussions and conclusions, respectively.

\section{Observation and data reduction}

\subsection{Spectroscopy}

Spectra of  $\beta$ Pic were obtained during 2010 December 16-18 (UT) with the low-resolution-10-$\micron$ mode (nominal resolution $R=116$ at 10 $\micron$) of T-ReCS. A 0.35$\arcsec$ slit was oriented along the disk at position angle 32$\degr$, the disk inclination measured by T5 using N- and Q-band images. Note that this value is very close to, but not exactly, the orientation of the secondary disk measured in the near-IR (32.7-33.8$\degr$, \citealt{lagrange2012}). The pixel scale was 0.09$\arcsec$. Eight integration sets of  $\beta$ Pic were acquired to give a total on-source time of 2,080 s, which was about 60\% of the time awarded for the program. The science observations were interlaced with those of the Cohen standard HD 39523 (K1 III, 5.1$\degr$ away from  $\beta$ Pic) in order to monitor the point spread function (PSF) and the flux/telluric correction. All data were taken in the standard chop-nod mode with a chop frequency of 2.3 Hz and a chop throw of 15$\arcsec$ perpendicular to the slit. 

These data were reduced with custom IDL routines. All frames in the raw T-ReCS FITS files were checked visually and $\sim$25$\%$ of them with low quality (e.g., bad seeing, strong background) were discarded before further reduction. Other artifacts, like non-flat background, channel offsets, jitter, and bad pixels, were corrected in each FITS file before combining the data.

The wavelength calibration was done using the telluric lines identified in the raw spectra. The flux and telluric corrections were performed using the Cohen model template spectrum of HD 39523 \citep{cohen1999}. We also calibrated the absolute flux at 11.7 $\micron$ using our imaging data, and these two methods agreed very well (within the typical mid-IR photometric accuracy of 10$\%$). 

One spectrum was acquired for each spatial pixel along the slit (corresponding to a 1.74-AU-wide region at the distance of $\beta$ Pic) within 20 AU to the star. As a double check of the photometric calibration accuracy, we integrated all reduced spectra within an aperture of 3.7$\arcsec$ centering on the star to mimic a single, low-spatial-resolution spectrum of $\beta$ Pic, and the result matched perfectly the data by \citet{knacke1993}.

In order to separate the disk excess from the total flux (star plus disk) at each wavelength, the spectral energy distribution (SED) of the stellar photosphere was assumed to be that of an 8,200 K blackbody with a flux density of 2.85 Jy at 8.0 $\micron$. This value was estimated by \citet{knacke1993} by matching the photospheric continuum to the near-IR photometric measurements of $\beta$ Pic. The uncertainty of their fit was not stated, but as we show in the following section, the uncertainty in the stellar flux should not affect the conclusions of our spectral analyses. The PSF profile at each wavelength was assumed to be Gaussian, the FWHM of which was 0.5$\arcsec$, measured from the spectrum of HD 39523.

\subsection{Imaging}

The two-epoch T-ReCS imaging data were obtained on 2003 December 30 (T5) and 2010 December 16 through the same Si-5 (11.7 $\micron$ ) filter. Key information about these two observations is summarized in Table~\ref{tbl-1}. Final images were co-added using frame registration to achieve the highest spatial resolution that the raw data could deliver. 

\begin{deluxetable}{lcc}
\tablewidth{0pt}
\setlength{\tabcolsep}{10pt}
\tablecaption{Summary of T-ReCS imaging of $\beta$ Pic\label{tbl-1}}
\tablehead{
\colhead{ } & \colhead{2003} & \colhead{2010}
}
\startdata
Program ID & GS-2003B-Q-14 & GS-2010B-Q-50 \\
UT Time & 2003 Dec 30 & 2010 Dec 16 \\
Integration Time (s) & $2\times450$ & $3\times300$ \\
Filter & \multicolumn{2}{c}{Si-5 (11.7 $\micron$)} \\
Pixel Scale (arcsec) & \multicolumn{2}{c}{0.09} \\
Mean Air Mass & 1.196 & 1.074 \\
FWHM\tablenotemark{a} (arcsec) & 0.36 & 0.35 \\
Instrument PA (degree) & 340 & 0 \\
Chop PA (degree) & 123 & 122 \\
\enddata
\tablenotetext{a}{Measured in the PSF references.}
\end{deluxetable}

To minimize the introduction of any spurious features due to row- or column-correlated pattern structure known to be associated with the detector array, T-ReCS was rotated so that the diskÕs major axis was tilted in the raw images. During the reduction, we rotated the data again to display the disk horizontally in the reduced images. 

Because our major science focus with the images is to examine the morphology, rather than the total flux, of the bright clump in the disk, we did not use the Cohen standard for the calibration. Instead, both images were normalized by the peak intensity of  $\beta$ Pic measured in each image through the Lorentzian profile fit. 

\section{Results and Analyses}

\subsection{Spectroscopic decomposition}
\label{sec:specanalysis}

\begin{figure}
\epsscale{1.0}
\plotone{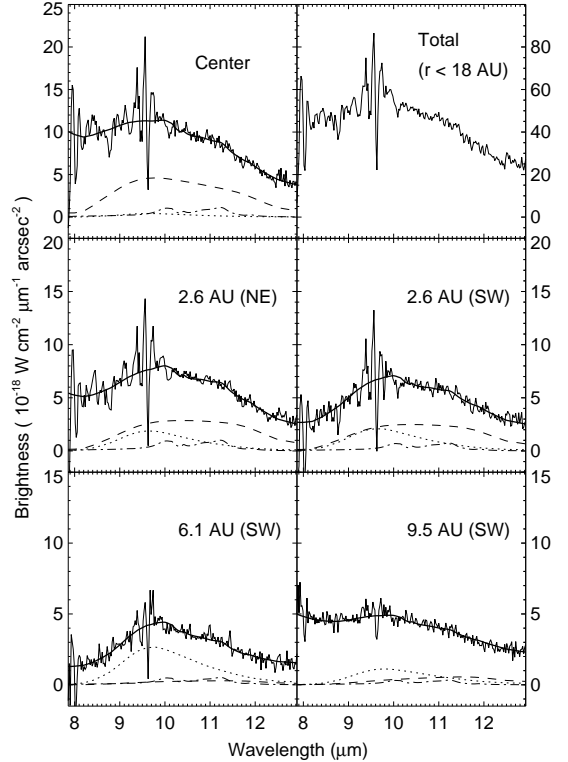}
\caption{Some example spectra of $\beta$ Pic (thin solid lines), overlapped with the best-fit spectral models (thick lines). Spectra are disk emission only, i.e., with stellar photospheric emission subtracted. Model spectra consist of four components: 2.0-$\micron$ glassy olivine (dashed), 0.1-$\micron$ glassy olivine (dotted), crystalline forsterite (dash-dotted), and power-law continuum (not shown). Error bars for each fitted component are shown in Figure~\ref{fig:profiles}.\label{fig:betapic_spectra}}
\end{figure}

The reduced N-band spectra (Figure~\ref{fig:betapic_spectra}) show clearly the broad 10-$\micron$ emission with a shape that depends on position across the inner disk of $\beta$ Pic. The shape of the 10-$\micron$ silicate feature, often quantified by the flux ratio of $F_{11.3}/F_{9.8}$, has long been considered a measure of the degree of dust processing \citep[dust growth and crystallization, e.g.,][]{bouwman2001, vanboekel2005, manoj2011}. Our spatially resolved spectra can be used to constrain the spatial distributions of amorphous and crystalline silicates across the disk of $\beta$ Pic, thus providing information on the processing and evolution of the dust. However, due to the strong telluric ozone feature around 9.8 $\micron$, obtaining a reliable estimate of the $F_{11.3}/F_{9.8}$ ratio is a challenge. In order to mitigate this problem, we followed an alternative approach developed by \citet{honda2003} and O4, and fit the entire 10-$\micron$ silicate feature with a spectral model composed of four emission components: 0.1-$\micron$ and 2.0-$\micron$ glassy olivine, sub-micron-sized crystalline forsterite, and a power-law continuum. The combination of the first two compounds was considered representative of a large size range (0.01-5 $\micron$) of amorphous silicates that are thought to dominate the 10-$\micron$ silicate feature \citep{bouwman2001}. The power-law continuum mimics all relatively featureless dust emission in the 10-$\micron$ window. The fitting formula is

\begin{equation}
\ I(\lambda )=a_{0}(\lambda /10\mu m)^{n}+\sum_{i=1}^{3}a_{i}\kappa _{i}^{\prime}(\lambda )B\left ( \lambda ,T_{i} \right )/B\left ( 10\mu m,T_{i} \right ).
\end{equation}

Here $I(\lambda)$ is the observed brightness (in $\mathrm{W\ cm^{-2}\ \micron^{-1}\ arcsec^{-2}}$) at wavelength $\lambda$ (in $\micron$), $a$ is the normalization factor, and $B$ is the Planck function. The parameter $n$ defines the slope of the power-law continuum. The subscript $i$ indicates four spectral components: continuum ($i=0$), 0.1-$\micron$ glassy olivine ($i=1$), 2.0-$\micron$ glassy olivine ($i=2$), and crystalline forsterite ($i=3$). The parameters $\kappa_i^{\prime}$ are the normalized mass absorption coefficients, and therefore are dimensionless. The normalized factor for each species is given by \citet{honda2003}. For the purpose of comparison, we chose to use the same data set of tabulated values of $\kappa$ as used in O4. For the 0.1- and 2.0-$\micron$ glassy olivine, $\kappa$ is calculated using Mie theory \citep{dorschner1995, honda2003}; for the crystalline forsterite, $\kappa$ is measured by \citet{koike2003}. $T$ is the dust temperature (in K). According to such a fitting formula, any remnant photospheric emission (due to inaccurate stellar subtraction) is not distinguishable from the underlying continuum, i.e., the first term in Equation 1, so it should not affect the fitting results of the other three dust components. 

As proposed by some studies, the mid-IR spectrum of $\beta$ Pic can be better reproduced by assuming core-mantle silicate particles (i.e., a cometary dust model), rather than the popular Draine and Lee interstellar silicates \citep[e.g.,][]{draine1984, li1998, chen2007}. However, we find that the latter fit our data fairly well, so we do not consider the core-mantled silicates in our work.

For amorphous olivine grains (i.e., $i=1,\ 2$), the temperature is derived by balancing the radiative absorption and re-emission (assuming spherical grains):

 \begin{equation}
 \int_{0}^{\infty }\kappa _{i}\left ( \lambda  \right )\left ( \frac{R_{*}}{r} \right )^{2}B_{\lambda }\left ( T_{*} \right )d\lambda =\int_{0}^{\infty }4\kappa _{i}\left ( \lambda  \right )B_{\lambda} \left ( T_{i}\left ( r \right ) \right )d\lambda 
 \end{equation}
 
where $R_*$ and $T_*$ are the stellar radius ($\mathrm{6.8\times10^{-3}}$ AU) and effective temperature (8,200 K) of $\beta$ Pic. The parameter $r$ is the disk radius, in AU, at the center of each spatial pixel. The temperature of the crystalline forsterite grains is assumed to be 1,000 K, i.e. approximately their annealing temperature. We note that, however, the fitted brightness distribution of forsterite is not sensitive to this assumption. Assuming a higher (e.g., 1,200 K, as in O4), lower, or varying temperature (i.e., similar to the small and big glassy olivine) affects the final results little.

For each spectrum associated with a certain spatial pixel, there are five free parameters, $a_i$ ($i=0,1,2,3$) and $n$, to fit the spectral model. The best fit was obtained by minimizing the reduced chi-square ($\chi^2_{\nu}$) over a reasonably large parameter space. The goodness of fit was also checked by eye, and we determined that all fits for which the $\chi^2_{\nu}$ deviated from the minimum $\chi^2_{\nu}$ by more than 10\% were visually distinguishable from the best-fit model, and were considered ``bad fits". In contrast, those for which the deviations were less than 10\% were considered ``good fits". For each free parameter, we then used all the values given by those ``good fits", and defined the standard deviation of that ensemble of values as our stated 1-$\sigma$ uncertainty for that parameter. 

Figure~\ref{fig:betapic_spectra} shows some examples of the model fit results. The data quality allows us to extract the 10-$\micron$ silicate feature out to about 20 AU ($\sim$1$\arcsec$) from the central star. Note that the ozone-dominated wavelengths (9.4-9.8 $\micron$) were not used in the evaluation of $\chi^2_{\nu}$. Although the peak around 11.3 $\micron$ does not appear to be particularly strong in the spectra, we found that excluding the crystalline forsterite component increases the $\chi^2_{\nu}$ by at least 20\% for the central pixels where the signal-to-noise ratios are high, and therefore we decided to keep this component in the spectral decomposition and analysis. 

\begin{figure}
\epsscale{1.0}
\plotone{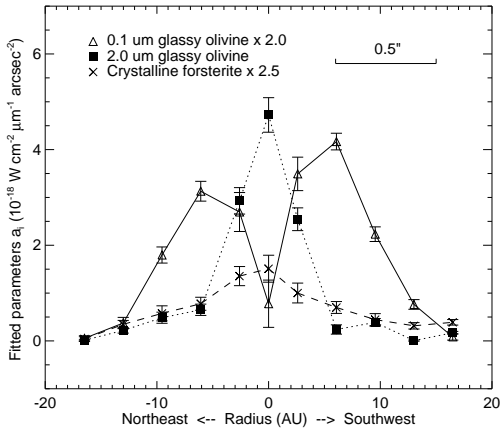}
\caption{Emission distributions of small (open triangles), big (filled squares), and crystalline grains (crosses) in the $\beta$ Pic disk. Note that the profiles are multiplied by different factors for the purpose of illustration. The approximate spatial resolution (FWHM) of T-ReCS at 10 $\micron$ is indicated in the top-right corner of the plot.\label{fig:profiles}}
\end{figure}

The spatial profiles of $a_1$, $a_2$, and $a_3$ are plotted in Figure~\ref{fig:profiles}. For a direct comparison with the result presented by O4, the spatial profiles are binned every two pixels (except for the central one) to closely match the spatial sampling used in O4. Figure~\ref{fig:profiles} shows clearly that the brightness distributions of three dust species are different from each other, as concluded by O4. The 2.0-$\micron$ glassy olivine particles (squares) have a single-peak distribution centered on the star; their brightness profile has a FWHM of $\sim$7 AU, comparable to the measured FWHM of the spectrum of the standard star ($\sim$9 AU at 10 $\micron$). In contrast, the 0.1-$\micron$ glassy olivine particles (triangles) show two almost symmetrically placed peaks, each at $\sim$5-6 AU from the star and $\sim$8 AU wide. The profile of crystalline forsterite particles (crosses) is weakly centrally peaked and symmetric within the uncertainties. We conclude that the crystalline forsterite particles are distributed across a region comparable in scale to that of the 2.0-$\micron$ glassy particles and unlike the double-peaked 2.0-$\micron$ glass olivine particles. 

\subsection{Imaging}

\begin{figure}
\epsscale{1.0}
\plotone{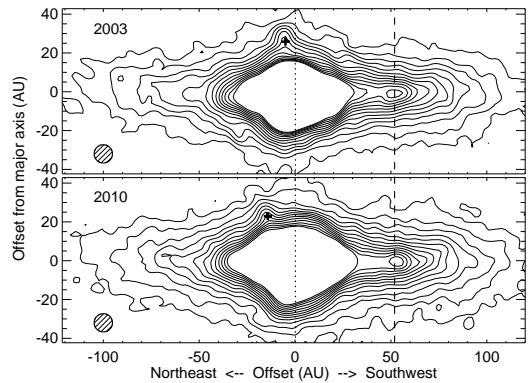}
\caption{Smoothed Si-5 (11.7 $\micron$) images of $\beta$ Pic, obtained in 2003 (upper panel) and 2010 (lower panel); NE to left, SW to right. The hatched circle in the lower left of each panel represents the spatial resolution achieved with these observations, being the smoothed FWHM of a point source. Both images have been rotated counterclockwise by 48$\degr$ (2003), or 68$\degr$ (2010). The vertical lines are at star (dotted) and at SW 52 AU (dashed). Contours are linearly spaced by 3-$\sigma$, starting at 3-$\sigma$. The features labeled by crosses are not real but ghost images of $\beta$ Pic. \label{fig:betapic_images}}
\end{figure}

Here we compare the mid-infrared images (smoothed with a Gaussian kernel of 3-pixel in FWHM) of $\beta$ Pic obtained in 2003 and 2010, focusing on the bright clump at 52 AU in the SW wing of the disk. For a comprehensive analysis of the 2003 images, per se, we refer the reader to T5. Two-epoch data are shown together in Figure~\ref{fig:betapic_images}. Note that some small morphological changes associated with the different instrument setups are evident. One of them is the ghost image of $\beta$ Pic (labeled by crosses), which rotated by $\sim20\degr$ with respect to $\beta$ Pic between two observations. This value is consistent with the change in instrument PA (Table~\ref{tbl-1}). Nevertheless, the two data sets were obtained under similar sky conditions, and their image qualities are comparable in terms of their FWHM and signal-to-noise ratios.

The bright clump, discovered by T5, embedded in the SW wing is unambiguously detected and resolved in both images. Note that the proper motion of $\beta$ Pic is $\sim0.5\arcsec$ (i.e., 9.6 AU at the distance of 19.28 pc, based on Hipparcos data, \citealt{vanleeuwen2007}) over seven years, a displacement that would have been obvious in the two-epoch images if the clump were a chance superposition of an unrelated background object; this observation confirms the conclusion of T5 that the clump is associated with the disk, and gives rise to the SW/NE asymmetry observed in the disk in earlier mid-IR observations \citep[e.g.,][]{pantin1997}.

\begin{figure}
\epsscale{1.0}
\plotone{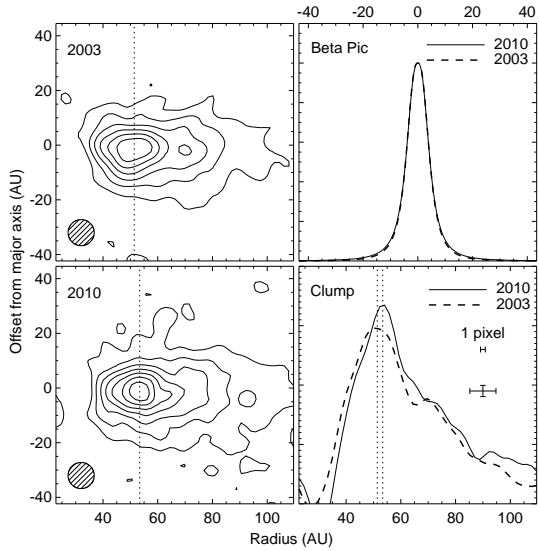}
\caption{The residual emission (left panels) of the SW 52 clump, resulting from subtracting the NE wing from the SW wing along a line through the star. Contours are spaced by 3-$\sigma$, starting at 3-$\sigma$. The profiles plotted in the lower-right panel show the varying flux (normalized by the peak intensity of $\beta$ Pic) along the major axis of the disk. The cross drawn at right shows the smoothed point source FWHM (horizontal bar) and the photometric error (vertical bar). The alignment uncertainty between the 2003 and 2010 profiles is estimated to be well below 1 pixel (i.e., 1.7 AU at the distance of $\beta$ Pic). The profiles drawn in the upper-right panel are best-fit Lorentzian PSF models of $\beta$ Pic obtained in the same data set.\label{fig:betapic_clump}}
\end{figure}

Aided by the line drawn at SW 52 AU, we can see that there seems to be a small offset between the 2003 and 2010 locations of the clump. As was done by T5 to investigate this region more carefully, we subtract the emission in the fainter NE wing from the SW wing, and draw the contours of the residual emission (assumed to be mainly contributed by the bright clump) in the left panels of Figure~\ref{fig:betapic_clump}. The profiles plotted in the lower-right panel show the flux distributions along the major axis of the disk. As a comparison, the best-fit Lorentzian profiles of $\beta$ Pic are also plotted in the upper-right panel, where the two-epoch data match each other almost perfectly; thus, there is no significant change in the PSF morphology between the 2003 and 2010 observations. Furthermore, if we divide each data set of 2003 or 2010 into two subsets differing in observing time and repeat the reduction procedures, the offset is still visible, which implies that changes in seeing are unlikely the cause of the offset.

To quantify the spatial displacement shown in the lower-right panel of Figure~\ref{fig:betapic_clump}, we performed a least-square fitting of the Lorentzian PSF model to the vicinity ($\sim$18 AU in width) around the clump in each image. The best-fit profiles suggested a radial shift of 2.0 AU between the 2003 and 2010 data. To estimate the error, we considered the following sources of uncertainties: (a) the centering error, i.e., how accurately the two-epoch data were aligned using the position of $\beta$ Pic; (b) uncertainty in the relative image orientations; (c) image distortion introduced by the image rotation algorithm. For (a) and (c), we performed a series of tests with artificial stars, and determined that under a typical noise level (S/N $\sim$1000), a 0.2-pixel centering accuracy is achievable; the distortion due to the image rotation algorithm is $\ll$0.1 pixel and thus negligible. We also checked the commissioning documents of T-ReCS, and found that the amount of image distortion is linearly proportional to the field of view. The maximum distortion is about 1 pixel, and  occurs at four corners of the detector array. Since the disk of $\beta$ Pic occupies only a small portion around the array center, and the relative image orientation between the 2003 and 2010 data is merely 20$\degr$, we estimated that the error due to (c) is no greater than 0.25 pixel. Taking all factors discussed here into account, the total uncertainty, which is the quadratic sum of (a), (b), and (c), is 0.32 pixel (i.e., 0.55 AU). Based on this, we conclude that the spatial displacement of the clump between 2003 and 2010 is marginally significant (3.6-$\sigma$) and therefore worth some consideration at this time. However, additional observations must confirm it over the coming years. We will discuss its scientific implications in the following section.

\section{Discussion}

\subsection{Radial distributions of dust grains}

\begin{figure}
\epsscale{1.0}
\plotone{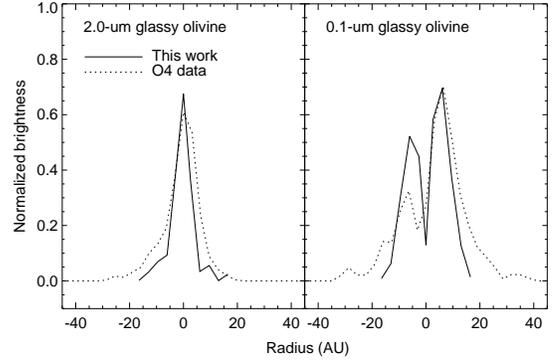}
\caption{Comparison between our work and the O4 results on the spatial distributions of the 0.1-$\micron$/2.0-$\micron$ glassy olivine. Note that different normalization factors are used for O4 and our data to aid visual comparison.\label{fig:comparison}}
\end{figure}

The spatial distributions of the three representative species derived by our spectral modeling have both similarities and important differences with those presented by O4. As seen in Figure~\ref{fig:comparison}, both studies show that the 2.0-$\micron$ glassy olivine is strongly peaked in the disk center, while the 0.1-$\micron$ glassy olivine does not show this concentration, but rather is double peaked. There is an apparent asymmetry in the brightness profile, and hence spatial distribution, of the 0.1-$\micron$ glassy olivine component. However, it is not as strong as that inferred by O4, and its significance not as firm. Also, we see a weaker crystalline feature: the total emission of forsterite (i.e., the profile shown in Figure~\ref{fig:profiles} integrated along the entire disk) is $\sim$20\% of that of the 2.0-$\micron$ glassy olivine, in contrast to $\sim$40\% in O4.

In the two studies (O4 and ours), the slits were placed along different PAs (30$\degr$ and 32$\degr$, respectively), but this is not sufficient to explain the disagreement. The slits used in both studies are $\sim$0.35" wide, thus not particularly narrow compared to the slit length of $\sim$1.5" (i.e., the diameter of the disk where the silicate feature is clearly detected). The offset due to the 2$\degr$ difference in PA is much smaller than the slit width, and thus should have little effect on the observations. Another difference between the two studies is the way the dust temperature is calculated. We found that the equation used by O4 is only appropriate for particles larger than typical wavelengths of absorbed stellar radiation, an assumption that is not valid for 0.1-$\micron$ grains, so we switched to another method as described in Section~\ref{sec:specanalysis}. However, we could not reproduce such a strong asymmetry, even if we adopted the same equation as in O4 to calculate the dust temperature.

We speculate that the different degrees of asymmetry observed for 0.1-$\micron$ grains by us and O4 can be explained by either a dynamical evolution of the inner disk of $\beta$ Pic during the time interval between our observations and those of O4, or an observational effect associated with the positioning of the slit by O4. At a radius of 5 AU, the dynamical timescale (i.e., the orbital period) in the disk of $\beta$ Pic is comparable to the time span between two observations (2003 and 2010, respectively). Thus, the change in the degree of asymmetry may be caused by the orbital movement of the parent dust-replenishing bodies that are not evenly distributed in the belt. However, O4 suggest that the asymmetry may be at least parly attributable to a slight de-centering, or shift, of the slit position toward the SW from optimum. In addition, the ozone feature between 9.4-9.8 $\micron$ can make the fit of the 0.1-$\micron$ glassy olivine problematic, especially at higher airmasses such as those ($>$3) which apply to the O4 data. Finally, the discrete spatial sampling associated with pixelation (i.e., the disk of $\beta$ Pic not being exactly symmetrically sampled by the pixels along the slit) may contribute to the observed asymmetries. This last effect is less important for our data, because the T-ReCS pixels are smaller. In the end, however, we conclude that there is currently no definitive way to account for the differences in our observations and those of O4.

\subsection{Crystalline forsterite}

As seen in Figure~\ref{fig:betapic_spectra}, the crystalline emission peak at 11.2 $\micron$ is not strong in our spectra (cf. Figure 1 of O4). The resultant spatial distribution of the crystalline forsterite is mildly centrally peaked, being several times broader at FWHM than that of the 2.0-$\micron$ glassy olivine. This result is consistent with the scenario of dust crystallization through thermal annealing by stellar radiation. In the disk of $\beta$ Pic, micron-sized grains can reach the classical annealing temperature of 1,000 K at $\sim$0.6 AU from the star. At such temperatures, the annealing timescale (of the order of hours to days, \citealt{hallenbeck1998}) is much shorter than the collisional timescale ($>10^4-10^5$ years, \citealt{ahmic2009}). In order to form an observable concentration, the newly formed crystalline grains must be orbitally stable, i.e., with a ratio of radiation pressure to gravity $\beta<0.5$. This holds true for grains larger than $\sim$1.3 $\micron$ \citep{artymowicz1997}. Much smaller (e.g., 0.1-$\micron$) grains can reach the annealing temperature at larger distances from the star, but are more affected by radiation pressure and repelled quickly (orbital timescales) from the star. For intermediate-sized grains, i.e., those with $\beta$ close to the blow-out limit, some degree of radial mixing following the thermal annealing is probable, and may account for the broadening in the spatial profile of forsterite relative to the 2.0-$\micron$ glassy particles. We conclude that, given the similar spatial distributions and characteristic particle sizes, it is possible that the crystalline forsterite grains are the annealed descendants of the larger glassy olivine particles.

Our admittedly speculative scenario clearly does not exclude other mechanisms that have been proposed to explain the formation of crystalline silicate minerals and their existence in the cold, outer disk. Those mechanisms, such as shock heating \citep{nuth2006, alexander2007}, aqueous alteration of dust on asteroidal parent bodies \citep{nuth2005}, and crystallization associated with the formation and subsequent destruction of massive fragments in young protostellar disks \citep{vorobyov2011}, can occur at large distances rather than the inner few AUs, but their signatures may not be observable by us.

\subsection{A planetesimal belt}
\label{sec:belt}

The double-peaked spatial profile of the 0.1-$\micron$ grains is interpreted by O4 as tracing a planetesimal belt that continuously replenishes the disk with these amorphous silicate grains. Due to the high $\beta$-value, sub-micron grains should be blown out of the disk in a very short timescale (less than several tens of years, \citealt{artymowicz1988}), and therefore the peaks of their spatial distribution should trace the dust-replenishment sites. Our observations imply that the radius of the belt proposed by O4 as the origin of the 0.1-$\micron$ grains is $5.6\pm3.0$ AU (assuming Gaussian profiles for each ``lobe"). If the planetesimal belt is an analogue of the asteroid belt in the Solar System (i.e., the belt is formed by the gravitational perturbation induced by a giant planet, and the inner and outer boundary of the belt are defined by the 4:1 and 2:1 resonance with the planet, \citealt{ferrazmello1994}), the measured location of the belt implies that the perturbing planet has an orbital radius of 9-14 AU. This is in rough agreement with the most up-to-date measurement of the orbital radius of $\beta$ Pic b (8-9 AU, \citealt{chauvin2012}). In addition, the orientation of the orbital plane of $\beta$ Pic b was recently determined to be $\sim$212$\degr$ \citep{lagrange2012}, i.e., in good alignment with the inner disk orientation measured in the mid-IR (T5). 

\subsection{An alternative interpretation for the different distributions of small/big grains}
\label{sec:nobelt}

As already noted, the brightness distribution of the 2.0-$\micron$ grains is very different from the double-peaked profile of the 0.1-$\micron$ particles. O4 account for this difference by proposing that the observed 2.0-$\micron$ grains, which they assume originate from the same planetesimal belt as the 0.1-$\micron$ particles, are propelled towards the central star by Poynting-Robertson drag, eventually concentrating near the disk center. In fact, this mechanism probably does not work in a disk as massive as $\beta$ Pic, where the collisional timescale is thought to be much shorter than that of the P-R drag (\citealt{artymowicz1997,wyatt2005}; see also \citealt{ahmic2009}). Rather, most of the grains do not travel far from their birthplaces before mutual collisions grind them fine enough to be expelled by radiation pressure. Therefore, we propose an alternative explanation for the difference in the brightness distributions of the two grain components, as follows.

\begin{figure}
\epsscale{1.0}
\plotone{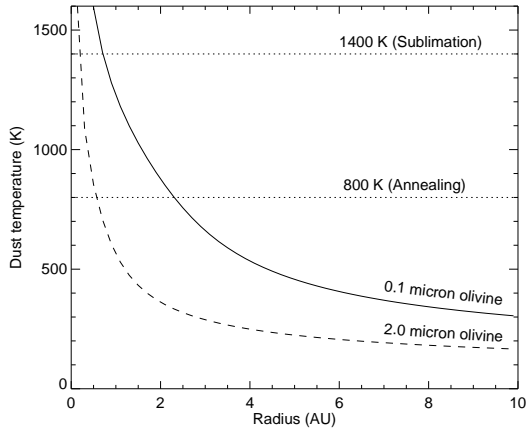}
\caption{Dust temperature calculated using Equation 2 for 0.1-$\micron$ and 2.0-$\micron$ glassy olivine, assuming a stellar luminosity of 8.7 $L_{\sun}$. Annealing and sublimation temperatures are adopted from \citealt{henning2009}.\label{fig:dust_temp}}
\end{figure}

Considering that small grains can reach a significantly higher temperature than do larger grains at the same radius of the disk (as shown in Figure~\ref{fig:dust_temp}, where temperature is calculated using Equation 2), the remarkable spatial profile of the 0.1-$\micron$ glassy olivine may be explained by a strong depletion of such grains around the disk center, as a consequence of thermal processing (annealing, sublimation) of the dust. That is, those particles at the center are removed or destroyed. According to Figure~\ref{fig:dust_temp}, 0.1-$\micron$ glassy olivine at $\sim$3 AU can be heated to 800 K, for which the annealing timescale is $\sim$ 3 yr \citep{fabian2000, tanaka2010}, i.e., comparable to the radiation pressure blow-out timescale (for sub-micron grains) and the orbital period at the radius of several AU, but significantly shorter than the collisional timescale of several thousand years \citep{artymowicz1997, ahmic2009}. Since the annealing and sublimation timescales decrease abruptly as temperature increases \citep{tanaka2010}, a deep depletion of 0.1-$\micron$ amorphous grains can be expected not far inside this radius. To reach the same temperature, 2.0-$\micron$ grains need to be as close as 0.6 AU to the star, and the central depletion zone would be too narrow to be detected by our observations. Thus, in this scenario, both dust species originate from parent dust populations that are centrally peaked. Their steady state distributions in the disk are modified by thermal annealing and sublimation, and coarsely sampled by our observations.

\begin{figure}
\epsscale{1.0}
\plotone{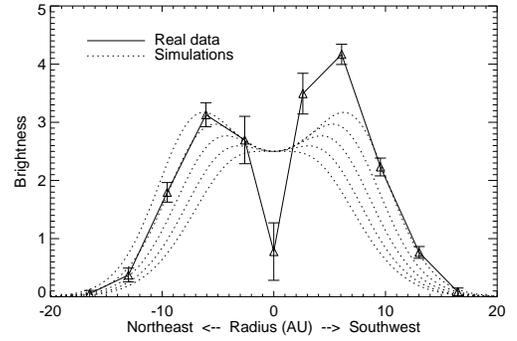}
\caption{Brightness profiles of the 0.1-$\micron$ glassy olivine, derived from real observations (solid line) or simulations (dotted lines), in which the $r_{in}$ (radius of the central clearing zone) are assumed to be (from bottom to top) 3, 4, 5, 6, and 7 AU, respectively. In all simulations, the spatial resolution is assume to be 6 AU.\label{fig:convol}}
\end{figure}

To see whether a central clearing zone of diameter 6 AU for the small amorphous grains is even resolvable by T-ReCS, we construct a toy model of the disk, in which the spatial distribution of 0.1-$\micron$ grains is Gaussian shaped but modified so that is has a central sharp-edged hole of radius $r_{in}$. This corresponds to a particle surface density peaked near $r_{in}$ and then decreasing gradually outwards. We also assume a uniform dust temperature and optically thin disk, so that the emission brightness is directly proportional to the amount of dust encountered by the line-of-sight. Finally, the intrinsic brightness profile is convolved with the PSF to mimic the real observations. In Figure~\ref{fig:convol}, the results are plotted against the observed spatial profile of the 0.1-$\micron$ glassy olivine. In all simulations, the FWHM of the PSF is set to be 7 AU, i.e., roughly the width of the brightness profile of 2.0-$\micron$ grains. We find that resolving a central clearing zone with $r_{in}$ of 3 AU would be a challenge for T-ReCS. Nevertheless, the hole would be clearly seen when $r_{in}$ is around 4-5 AU, close to the value of $r_{in}$ suggested by the dust annealing temperature as discussed above. We also note that our model is unable to reproduce the great depth of the central dip, which is very noticeable in the real data but not so in the simulations. This may indicate that the seeing was significantly better during the observations of $\beta$ Pic than when the PSF was observed, or it may reflect additional uncertainties in the photospheric subtraction or the spectral fitting.

\subsection{The dust clump at SW 52 AU}

Using multi-wavelength mid-IR images, T5 explored the properties of the clump at SW 52 AU. The SED of the clump implies a characteristic particle temperature there of $\sim$190 K, noticeably higher than the value of about 140 K for the region on the opposite side of the disk, at NE 52 AU. Therefore, T5 infer that a population of sub-micron-sized (0.1-0.2 $\micron$) particles exists in the clump, with their sizes and/or compositions different from the dust elsewhere in the disk. They conclude that the catastrophic collisional break-up of a planetesimal ($\sim$100 km or larger in diameter) is a plausible explanation for this localized inhomogeneity in the dust population. In this picture, fragment post-collision velocities of $\sim$1-2 $\mathrm{km\ s^{-1}}$ and radiation-pressure-driven dispersal account for the vertical and radial extent of the clump. Several studies \citep[e.g.,][]{dermott2002, kenyon2004, grigorieva2007} indicate that such catastrophic collisions should indeed be observable. 

\begin{figure}
\epsscale{1.0}
\plotone{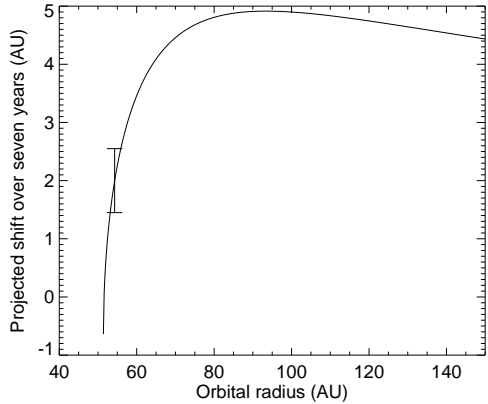}
\caption{The projected displacement expected for an object orbiting $\beta$ Pic in a circular, edge-on Keplerian orbit over 7 years, assuming a stellar mass of 1.75 $M_{\sun}$. The vertical error bar indicates the observed radial displacement and its uncertainty ($2.0\pm0.6$ AU), corresponding to the most probable orbital radius of $54.3^{+2.0}_{-1.2}$ AU.\label{fig:clump_offset}}
\end{figure}

The two images taken seven years apart may shed additional light on the clump properties. Figure~\ref{fig:clump_offset} shows the projected displacement over seven years expected for an object orbiting $\beta$ Pic in an edge-on, circular Keplerian orbit. If we assume that the orbital movement of the clump is traced by the locations of the peaks in the emission residuals shown in Figure~\ref{fig:betapic_clump}, then the measured projected shift of 2.0 AU between 2003 and 2010 is consistent with a true orbital radius of $54.3^{+2.0}_{-1.2}$ AU. Since the SW disk is rotating toward us, as suggested by high resolution Na I D spectroscopy of $\beta$ Pic \citep{olofsson2001}, the observed shift also implies that clump is moving towards us, and approaching its greatest southwestern elongation. This tentative conclusion can be tested over the coming years, and, if confirmed, would constitute unique insight into the dynamics of a debris disk; to date, $\epsilon$ Eri is the only other debris disk for which orbital proper motions may have been measured \citep{greaves2005}.  

Also note that some features of the dust clump, like a high proportion of sub-micron grains, physical extensions in both vertical and radial directions, and the orbital movement, are in general consistent with the consequence of a collisional avalanche, i.e., a chain reaction initiated by a break-up of a comet or a planetesimal-belt object, producing a substantial amount of small grains propagating outwards in the debris disk \citep{artymowicz1996, grigorieva2007}. Although in this scenario, the current morphology of the clump indicates that we are witnessing a rare event only several tens to several hundred year after the catastrophic break-up, such a possibility cannot be ruled out. Current data are not sensitive enough to allow us to explore the grain properties in this region as we have done for the innermost part of the disk. Deeper spectroscopic observations in the future will be crucial to understand the nature of this striking feature in the disk of $\beta$ Pic.

Finally, we note that, although the clump is clearly seen in the mid-IR, no counterpart has been found at optical or near-IR wavelengths \citep{golimowski2006, boccaletti2009}. This is not surprising, since the scattered light and mid-IR emission may trace different regions of the disk. Nevertheless, this issue has not been studied in detail, and important hints of the dusty environment and formation mechanism of the clump can be revealed by a model that is able to reproduce the multi-wavelength observations simultaneously. 

\section{Summary}

We observed the edge-on, planet-bearing disk of $\beta$ Pic using T-ReCS at Gemini to clarify and extend previous observations and conclusions about this unique system. We used spectroscopy and spectral modeling of the 10-$\micron$ silicate feature to constrain the spatial distributions of three representative dust components (0.1-$\micron$/2.0-$\micron$ glassy olivine and crystalline forsterite) across the inner 20-AU of the disk. We also obtained N-band images of the $\beta$-Pic disk. We conclude the following:

1. We confirm the O4 results that the 2.0-$\micron$ glassy olivine is strongly peaked in the disk center and that, in contrast, the 0.1-$\micron$ glassy olivine does not show this concentration, but rather is double peaked, with the peaks on either side of the star. However, we did not see the strong brightness asymmetry between those two peaks of the 0.1-$\micron$ glassy olivine reported by O4. 

2. Compared to the observations by O4, the distribution of the crystalline silicates observed by us is much weaker and not so strongly centrally peaked.
 
3. Although the spatial distribution of the 0.1-$\micron$ glassy olivine is consistent with the general scenario of a dust-replenishing planetesimal belt embedded in the disk, as proposed by O4, we draw attention to an alternative interpretation in which no planetesimal belt is required, but rather some of the disk structure is associated with grain destruction or alteration by sublimation or annealing. 
 
4. By comparing the imaging data with that acquired in 2003, we confirmed the existence of the dusty clump at 52 AU in the SW disk. We speculate that the clumpÕs projected spatial displacement of $\sim$2.0 AU, a 3.6-$\sigma$ result, between two epochs separated by seven years may be due to the Keplerian motion of the clump at an orbital radius of $54.3^{+2.0}_{-1.2}$ AU.

\acknowledgments
This study is based on observations obtained at the Gemini Observatory through the program GS-2003B-Q-14 and GS-2010B-Q-50. Gemini observatory is operated by the Association of Universities for Research in Astronomy, Inc., under a cooperative agreement with the NSF on behalf of the Gemini partnership: the National Science Foundation (United States), the Science and Technology Facilities Council (United Kingdom), the National Research Council (Canada), CONICYT (Chile), the Australian Research Council (Australia), Minist\'erio da Ci\.encia e Tecnologia (Brazil) and Ministerio de Ciencia, Tecnolog\'ia e Innovaci\'on Productiva (Argentina). The authors thank Mitsuhiko Honda for kindly providing the dust mass absorption coefficient data, and the anonymous referee for very useful comments and suggestions which have greatly improved the quality and clarity of this paper. CMT gratefully acknowledges NSF support for this work through awards AST-0903672 and AST-0908624. CMW acknowledges support from the Australian Research Council through the Future Fellowship grant FT100100495.

Facilities: \facility{Gemini:South(T-ReCS)}

\clearpage

\bibliographystyle{apj}
\bibliography{apj-jour,references}
  
\end{document}